\documentclass[preprint,nofootinbib]{revtex4}

\usepackage{graphicx}
\usepackage[usenames]{color}

\begin{document}



\title{Density perturbations in decaying holographic dark
energy models}

\author{Kyoung Yee Kim, Hyung Won Lee and Yun Soo Myung}
\email{ysmyung@inje.ac.kr} \affiliation{Institute of Basic Science
and School of Computer Aided Science\\Inje University, Gimhae
621-749, Korea}


\begin{abstract}
We study cosmological perturbations in the context of an
interacting dark energy model, where the holographic dark energy
with IR cutoff decays into the cold dark  matter (CDM). For this
purpose, we introduce three IR cutoffs of Hubble horizon, particle
horizon, and future event horizon.   Here we present  small
perturbations under the case that effective equation of state
(EOS: $\omega^{\rm eff}$) for the holographic energy density is
determined to be the same negative constant  as that for the CDM.
Such new matter  productions tend to dilute the density
perturbations of CDM (matter contrast). For a decelerating
universe of $\omega^{\rm eff}>-1/3$, the matter contrast is
growing as the universe evolves, while for an accelerating
universe of $\omega^{\rm eff}<-1/3$, the matter contrast is
decaying, irrespective of the choice of IR cutoff. This shows
clearly that the acceleration suppresses the growing of the
density perturbations at the early universe.
\end{abstract}

\maketitle

\section{Introduction}

The cosmological constant problem has acquired a renewed
importance since several independent observations have been
pointing to the presence of a negative pressure component in the
cosmic fluid~\cite{Pad}. In  view of quantum field theories, the
natural candidate for such a dark energy is the quantum vacuum
energy. Since it has the symmetry of the background, its
energy-momentum tensor has the form $T_{\mu\nu}= \Lambda
g_{\mu\nu}$, where $\Lambda$ is a scalar function of coordinates.
This leads to the equation of state $p_{\Lambda} = -
\rho_{\Lambda} = - \Lambda$, where $\Lambda$ may be a function of
time ($\Lambda(t)$) in general. In the case of a constant
$\Lambda$, the vacuum contribution plays the role of a
cosmological constant in Einstein's equations. The model based on
a constant ${\rm \Lambda}$ and a variable ${\rm \Lambda}(t)$ is
called the ${\rm \Lambda}$CDM model and  ${\rm \Lambda}(t)$CDM
model, respectively.

The idea of a time-dependent cosmological term has provided
different phenomenological implementations ~\cite{Ozer}, being a
subject of renewed interest in recent years~\cite{dim,wm,Fabris}.
A general feature of all those approaches is the production of a
new kind of matter, compatible  with the vacuum decay in order to
assure the conservation of the total
energy-momentum~\cite{Barrow}.

On the other hand, there exists  dynamical cosmological constants
derived from the holographic principle.  Cohen {\it et al} have
shown that in quantum field theory, the UV cutoff
$\tilde{\Lambda}$ is related to the IR cutoff $L_{\rm \Lambda}$
due to the limit set by forming a black hole~\cite{CKN}. In other
words, if $\rho_{\rm \Lambda}=\tilde{\Lambda}^4$ is the quantum
vacuum energy density caused by the UV cutoff, the total energy of
system with size $L_{\rm \Lambda}$ should not exceed the mass of
the system-size black hole: $L_{\rm \Lambda}^3 \rho_{\rm
\Lambda}\le 2L_{\rm \Lambda}M_p^2$.  If the largest  size $L_{\rm
\Lambda}$ is chosen to be the one saturating this bound,  the
holographic energy density is then given by $\rho_{\rm \Lambda}=
3c^2M_p^2/8\pi L_{\rm \Lambda}^2$ with a parameter
$c$~\cite{HSU,LI,Myung2}, in contrast with the conventional energy
density of $\rho \propto 1/L^3$. We consider $\rho_{\rm \Lambda}$
as the dynamical cosmological constant.

Also, we have two different views of determining the equation of
state for the holographic energy density.  The first view is that
its native equation of state is not changing as the universe
evolves~\cite{HOV}. It is fixed by $\omega_{\rm \Lambda 0}=-1$
initially. An important point to note is that the holographic
energy density itself is changing as a result of decaying into the
CDM. According to the total energy-momentum conservation, its
change must be compensated by the corresponding change in the  CDM
sector~\cite{Zim3}.  The second view is that the equation of state
for the holographic energy density is not fixed in general, but it
is changing as the universe expands with/without the
interaction~\cite{HSU,LI}. If an interaction is present between
two matters, one would be better to use the effective equation
 of state  $\omega^{\rm eff}_{\rm \Lambda}$~\cite{KLM}
 than the native equation of state $\omega_{\rm \Lambda}$~\cite{WGA}.
In the presence of an interaction, the first view  leads to a
constant EOS $\omega^{\rm eff}_{\rm \Lambda}$, while the second
view implies a dynamical EOS $\omega^{\rm eff}_{\rm
\Lambda}(\Omega)$ with density parameter $\Omega_i=\rho_i/\rho_c$.

In this work, we investigate the vacuum decaying  of holographic
energy density with the IR cutoff using the first view of the
constant EOS. The key of our system is an interaction between
holographic energy and CDM. They are changing as a result of
energy transfer from holographic energy to the CDM. We call this
the H($t$)CDM model, similar to the ${\rm \Lambda}(t)$CDM
model~\cite{wm}. Specifically, we present small perturbations
under the case of a new matter production with $\omega^{\rm
eff}_{\rm m}=\omega^{\rm eff}_{\rm \Lambda}$. In general, such a
matter production may tend to dilute the matter contrast. For a
decelerating universe of $\omega^{\rm eff}_{\rm m}>-1/3$, the
matter contrast is growing as the universe evolves, while for an
accelerating universe of $\omega^{\rm eff}_{\rm m}<-1/3$, the
matter contrast is decaying, irrespective of the choice of IR
cutoff. This shows the connection between the background evolution
and matter contrast clearly.

\section{The model}
For a flat universe composed of cold dark matter and holographic
energy density~\cite{Myungs}, the first Friedmann equation is
given by
\begin{equation} \label{FFEQ}
H^2=\frac{1}{3 m_p^2}(\rho_\Lambda+\rho_m)
\end{equation}
where  $H=\dot{a}/a$, $m_p=1/\sqrt{8\pi G}$ is the reduced Planck
mass, and the holographic dark energy takes the form
\begin{equation}
\rho_\Lambda=\frac{3c^2m_p^2}{L_\Lambda^2}.
\end{equation}
The conservation law of the total energy-momentum leads to
\begin{equation}
\dot{\rho}_T+3H(\rho_T+p_T)=0,
\end{equation}
where the overdot($\dot{}$) denotes the derivative with respect to
the cosmic time $t$. Here $\rho_T=\rho_m+\rho_{\rm \Lambda}$ and
$p_T=p_m+p_{\rm \Lambda}$ with $p_m=0$ and $p_{\rm
\Lambda}=-\rho_{\rm \Lambda}$. We are interested in an interacting
case with $q\not=0$. In this case, the conservation law  is split
into two equations
\begin{equation} \label{conl}
\dot{\rho}_m + 3H(\rho_m + p_m)=q, ~~~~ \dot{\rho}_{\rm
\Lambda}=-q.
\end{equation}
For a choice of the interaction $q=\epsilon H \rho_m$~\cite{KLMM},
the solution to the above equations is given by
\begin{equation} \label{DEN}
\rho_m=\rho_{m0} a^{-3(1-\epsilon/3)}, ~~~
\rho_\Lambda=\frac{\epsilon}{3-\epsilon}\rho_m.
\end{equation}
Then, we obtain an explicit form of the solution:
\begin{equation}
a(t) =  \left[
\frac{\rho_{m0}}{4m_p^2}t^2\right]^{\frac{2}{3-\epsilon}},~H=
\frac{\dot{a}}{a} = \frac{2}{3-\epsilon}\frac{1}{t}.
\end{equation}
Their energy densities take the forms
\begin{equation}
\rho_m=\left[\frac{4 m_p^2}{3-\epsilon}\right]
\frac{1}{t^2},~\rho_\Lambda=\left[\frac{4 m_p^2
\epsilon}{(3-\epsilon)^2}\right] \frac{1}{t^2}.
\end{equation}
Finally, the constant effective EOS could be read off from
Eq.(\ref{DEN}) as
\begin{equation}
\omega_{\rm m}^{\rm eff}=-\frac{\epsilon}{3}=\omega_{\rm \Lambda}^{\rm eff}
\end{equation}
which depends on the choice of IR cutoff. As will be shown in Fig.
3, $\omega_{\rm m}^{\rm eff}(c)$ depends critically  on the choice
of IR cutoff $L_{\rm \Lambda}$.

On the other hand, the Newtonian equation governing the evolution
of density perturbations of CDM~\cite{MFB,BRAN} can be generalized
in order to account for matter production. It is given
by~\cite{Waga}
\begin{equation} \label{Waga}
\ddot{\delta} + \left(2H + \frac{q}{\rho_m}\right) \dot{\delta} -
\left[ \frac{\rho_m}{2 m^2_p} - 2H\frac{q}{\rho_m} -
\dot{\left(\frac{q}{\rho_m}\right)}\right] \delta = 0,
\end{equation}
where $\delta = \delta \rho_m/\rho_m$ is the density perturbations
of the cold dark matter (matter contrast) and $q$ is the source of
matter production defined by Eq.(\ref{conl}).  In the case of a
constant $\rho_{\rm \Lambda}=\tilde{\Lambda}^4(q = 0)$,
(\ref{Waga}) reduces to the usual non-relativistic equation for
the linear evolution of the matter contrast. That is, it
corresponds to the matter contrast of the ${\rm \Lambda}$CDM
model. In our case of the  H($t$)CDM model, we have  $q = -
\dot{\rho}_{\rm \Lambda} =\epsilon H \rho_m$. Finally, the
Newtonian equation of density perturbations with the holographic
dark energy is given by
\begin{equation} \label{denseq}
\ddot{\delta}+ \left[ \frac{4+2 \epsilon}{3- \epsilon}\right]
\frac{\dot{\delta}}{t} -\left[ \frac{2}{3- \epsilon}-\frac{8
\epsilon}{(3- \epsilon)^2} + \frac{2 \epsilon}{3-
\epsilon}\right]\frac{\delta}{t^2}=0.
\end{equation}
At this stage, we point out the difference between our model and
decaying vacuum cosmology~\cite{Borges}. The latter case has
considered the $\Lambda(t)$CDM model with a different interaction
$q=\sigma H$.

\section{Non-interacting case: ${\rm \Lambda}$CDM model}

\begin{figure}[b]
\begin{center}
\includegraphics[scale=0.9]{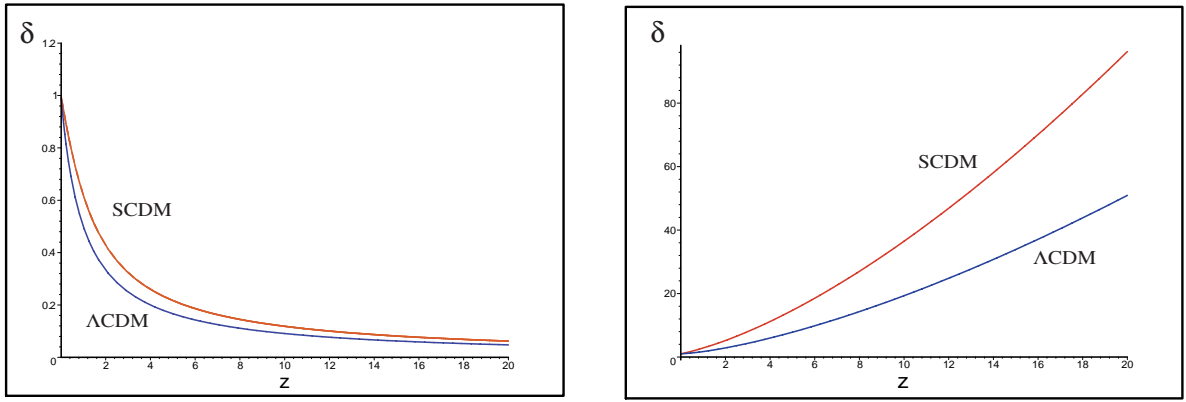}
\end{center}
\caption{\footnotesize The density contrasts as a function of the
redshift $z$ for $\Lambda$CDM and SCDM.  The left panel represents
for growing modes, while the right one denotes the decaying modes.
}
\end{figure}

First of all, we discuss the non-interacting holographic dark
energy model with $q=0$ in Eq.(\ref{conl}), which is identical to
the ${\rm \Lambda}$CDM model. In this case, we have the EOS of
$\omega_{\rm \Lambda }=-1$ for the holographic energy density like
the cosmological constant and $\omega_{\rm m }=0$ for the CDM.

The evolution equation for the density perturbation is given by
\begin{equation}
\delta'' +\Bigg[ \frac{1}{2} - \frac{3}{2} \Omega_{\rm \Lambda}(x)
\Bigg] \delta' -\frac{3}{2} \Bigg[ 1 - \Omega_{\rm \Lambda}(x)
\Bigg] \delta = 0, \label{evol_pert}
\end{equation}
where $'$ denotes derivative with respect to  $x = \ln a$.    The
density parameter  $\Omega_{\rm \Lambda}(x)$ satisfies the
background evolution equation,
\begin{equation}
\Omega_{\rm \Lambda}' = 3 \Omega_{\rm \Lambda} ( 1 - \Omega_{\rm
\Lambda} ), \label{back_evol}
\end{equation}
while the first Friedmann equation (\ref{FFEQ}) is $\Omega_{\rm
\Lambda}+\Omega_{\rm m}=1$. The solution to this equation is given
by
\begin{equation}
\Omega_{\rm \Lambda}(x) = \frac{18}{7}
\frac{e^{3x}}{1+\frac{18}{7}e^{3x}}, \label{Omega_Lambda}
\end{equation}
where the numerical factor $\frac{18}{7}$ is chosen to fit the
present condition of $\Omega_{\rm \Lambda}(0) = 0.72$ and
$\Omega_{\rm m}(0) = 0.28$.

If there is no cosmological constant term, then
Eq.(\ref{evol_pert}) becomes
\begin{equation}
\delta'' + \frac{1}{2} \delta' -\frac{3}{2}  \delta = 0,
\label{evolv_cdm}
\end{equation}
which describes the matter contrast of the standard cold dark
matter (SCDM) with $\omega_{\rm m}=0$.  The solution for this
equation is given by
\begin{equation}
\delta(x) = C_1 e^x + C_2 e^{-3x/2}  \label{sol_cdm}
\end{equation}
with two constants $C_1$ and $C_2$. Obviously, the first term  is
a growing mode and the latter is a decaying one. This could be
easily conjectured because of the purely decelerating phase of the
CDM.

On the other hand, the general solution of $\Lambda$CDM to
Eq.(\ref{evol_pert}) is given by
\begin{equation}
\delta(x) = C_1 \frac{1}{\sqrt{\Omega_{\rm \Lambda}(x)}}
\int_{-\infty}^x \Omega_{\rm \Lambda}^{3/2}(y) e^{-2y} dy +
           C_2\frac{1}{\sqrt{\Omega_{\rm \Lambda}(x)}},
\label{sol_lcdm}
\end{equation}
where $\Omega_{\rm \Lambda}(x)$ is given by
Eq.(\ref{Omega_Lambda}). The connection to the redshift $z$ is
given by $x=-\ln(1+z)$ with $a=1/(1+z)$. The first term
corresponds to a growing solution, while the second is a decaying
one. In this case, we obtain  a growing mode even the universe is
composed of the cosmological constant with $\omega_{\rm
\Lambda}=-1$ and the CDM with $\omega_{\rm m}=0$. This is possible
because the early universe is the CDM-dominated phase. However, as
is shown in Fig. 1, the growing rates for SCDM and $\Lambda$CDM
are different. The growing rate for SCDM is greater than that of
$\Lambda$CDM since the CDM-nature of $\Lambda$CDM  decreases as
the universe evolves (the universe becomes the dark
energy-dominated phase).

\section{H($t$)CDM with Hubble horizon}

\begin{figure}[b]
\begin{center}
\includegraphics[scale=0.9]{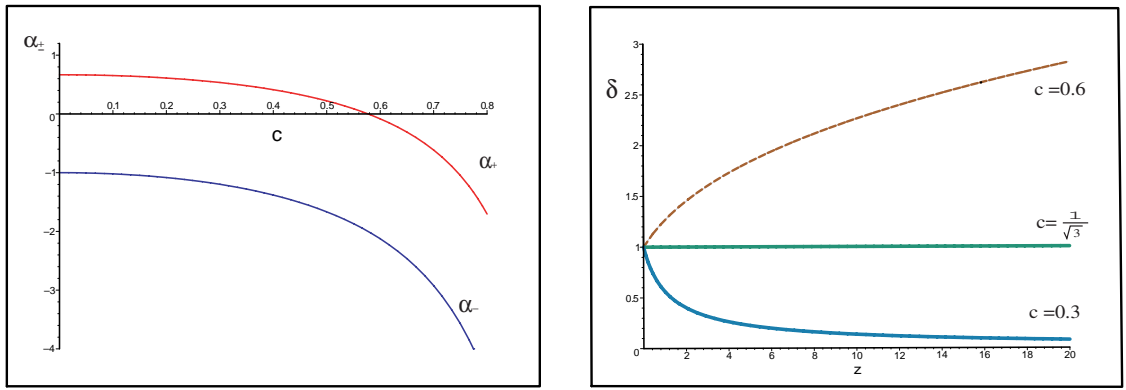}
\end{center}
\caption{\footnotesize The matter contrast as a function of the
redshift $z$ for H($t$)CDM with Hubble horizon. The left panel
denotes the exponents $\alpha_{\pm}$ and the right one denotes the
matter contrasts for $\alpha_+$ with different
$c=0.6,1/\sqrt{3},0.3$ from top to bottom. }
\end{figure}
We choose the IR cutoff as Hubble horizon with
$L_\Lambda=R_{HH}=1/H$. In this case, one has
$\epsilon=3c^2$~\cite{Myungs}. The evolution equation takes the
form
\begin{equation} 2 \dot{H}+3(1-c^2)H^2=0.
\end{equation}
 The corresponding solution
is given by \cite{Borges}
\begin{equation} \label{a}
~H = \frac{2}{C_1 + 3  (1-c^2) t},a(t) = C_2 \left[C_1 +
3(1-c^2)t\right]^{\frac{2}{3(1-c^2)}},
\end{equation}
where  $C_1$ and $C_2$ are  integration constants. Assuming the
initial condition of  $a=0$ at $t=0$, we obtain
\begin{equation}
a(t) = C_2 \left[ 3(1-c^2)t\right]^{\frac{2}{3(1-c^2)}},~H =
\frac{2}{3 (1-c^2)}\frac{1}{t}.
\end{equation}
The energy densities are given by
\begin{equation}
\rho_\Lambda=3 c^2 m_p^2 H^2 =\frac{4 c^2 m_p^2}{3 (1-c^2)}
\frac{1}{t^2},~\rho_m=\frac{4 m_p^2}{3} \frac{1}{t^2}.
\end{equation}
We check that the ratio of two energy densities is fixed as
\begin{equation}
\frac{\rho_m}{\rho_\Lambda}=\frac{1-c^2}{c^2}.
\end{equation}
Finally, the effective EOS is given by
 \begin{equation}
\omega_{\rm m}^{\rm eff}=-\frac{\epsilon}{3}=-c^2=\omega_{\rm
\Lambda}^{\rm eff}.
\end{equation}
An important point to note is that the interaction between two
matters  is a mechanism to generate the  matter production.   In
the case of a homogeneous production, the new matter tends to
dilute the density perturbations, leading to a suppression of the
density contrast. Plugging $\epsilon=3c^2$ into Eq.(\ref{denseq}),
the Newtonian equation for matter contrast is
\begin{equation}
\ddot{\delta}+ \left[ \frac{4+6 c^2 }{3(1-c^2)}\right]
\frac{\dot{\delta}}{t} -\left[ \frac{2}{3 (1-c^2)}-\frac{8 c^2}{3
(1-c^2)^2} + \frac{2 c^2}{(1-c^2)}\right]\frac{\delta}{t^2}=0
\end{equation}
In order to solve the above equation,  we assume that
$\delta(t)=t^\alpha$. Then its exponents are given by
\begin{equation}
\alpha_\pm(c)=\frac{1}{2}\left[\frac{3 (1+c^2)+4}{3
(1-c^2)}\right] \pm \sqrt{\left(\frac{4+3  (9c^2-1)}{6
(1-c^2)}\right)^2+ \frac{2-2c^2(3c^2+1)}{3(1-c^2)^2}}.
\end{equation}
Fig. 2 shows that $\alpha_+$ is positive, zero, and negative  for
$c<,=,>1/\sqrt{3}$, respectively, while $\alpha_-$ is always
negative. Hence we find a growing mode of the matter contrast only
for $0<c<1/\sqrt{3}$. This means that the decelerating phase leads
to a growing mode, while the accelerating phase of $c>1/\sqrt{3}$
implies a  decaying mode. Finally, $c=1/\sqrt{3}$ provides a
constant mode.

\section{H($t$)CDM with particle horizon and future event horizons}
Choosing the particle horizon (PH)  and  future event  horizon
(FH) leads to~\cite{Myungs}
\begin{equation}
\epsilon_{\rm PH/FH}=1+\frac{2}{3c^2}\mp
\frac{2\sqrt{3c^2+1}}{3c^2}.
\end{equation}
The effective EOS are given by
\begin{equation}
\omega_{\rm m}^{\rm eff}=\omega_{\rm PH/FH}^{\rm eff}=-\frac{\epsilon_{\rm
PH/FH}}{3}.
\end{equation}
The Newtonian equations take the forms
\begin{equation}
\ddot{\delta}_{\rm PH/FH}+ A_{\rm PH/FH}\frac{\dot{\delta}_{\rm
PH/FH}}{t}-B_{\rm PH/FH}\frac{\delta_{\rm PH/FH}}{t^2}=0
\end{equation}
with
\begin{equation}
A_{\rm PH/FH}= \frac{4+2 \epsilon_{\rm PH/FH}}{3- \epsilon_{\rm
PH/FH}},~B_{\rm PH/FH}= \frac{2}{3- \epsilon_{\rm PH/FH}}-\frac{8
\epsilon_{\rm PH/FH}}{(3- \epsilon_{\rm PH/FH})^2} + \frac{2
\epsilon_{\rm PH/FH}}{3- \epsilon_{\rm PH/FH}}.
\end{equation}
 Assuming that $\delta_{\rm PH/FH}=t^{\alpha^{\pm}_{\rm
PH/FH}}$, we could obtain its exponents
\begin{equation}
\alpha^{\pm}_{\rm PH/FH}=\frac{1-A_{\rm PH/FH}}{2}\pm
\frac{\sqrt{(A_{\rm PH/FH}+1)^2+B_{\rm PH/FH}}}{2}.
\end{equation}
We show its behavior in graphically because its forms are very
complicated. As is shown in Fig. 3, the effective EOS with the
particle horizon has a bound of $-1/3 \le \omega_{\rm m}^{\rm eff}
\le 0$, which means that this case always provides the
decelerating universe for any $c$. Hence we find from Fig. 4 that
the growing modes $\delta_{\rm PH}$ appears for any $c$ when
choosing $\alpha_+$.
\begin{figure}[h]
\begin{center}
\includegraphics{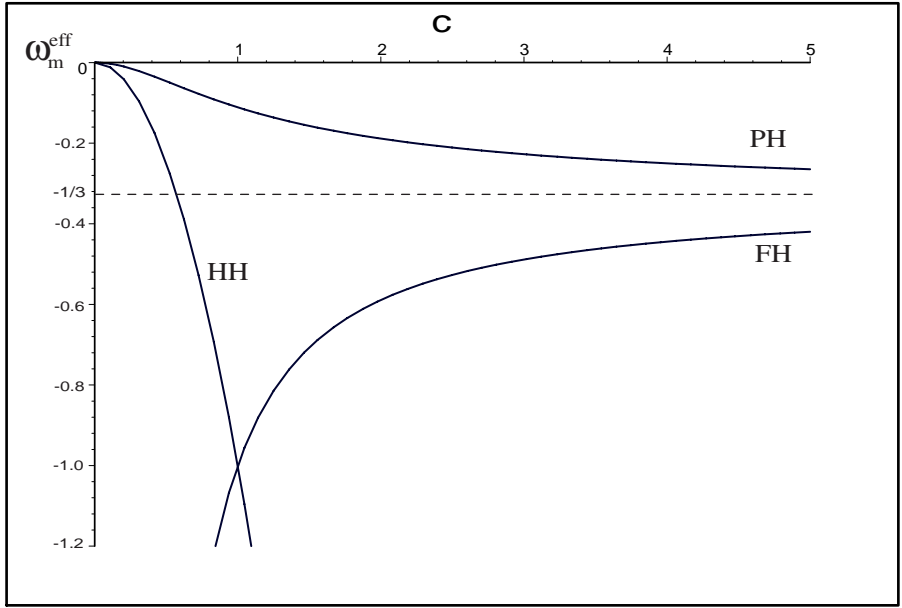}
\end{center}
\caption{\footnotesize The three effective EOS $\omega_{\rm
m}^{\rm eff}$ as a function of the parameter $c$ for H($t$)CDM
with Hubble horizon (HH), particle horizon  (PH), and future event
horizon (FH), respectively. In the case with FH, we have a region
of $0< c<1$, which may represent the phantom phase of $\omega^{\rm
eff}_{\rm m}<-1$. }
\end{figure}

\begin{figure}
\begin{center}
\includegraphics[scale=0.9]{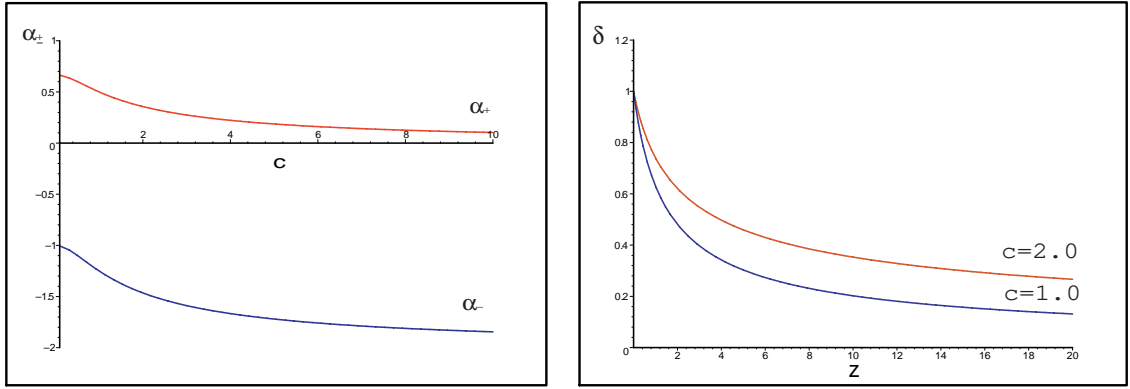}
\caption{\footnotesize The matter contrast as a function of the
redshift for H($t$)CDM with particle horizon (PH). The left panel
denotes the exponents $\alpha_{\pm}$  and the right one denotes
the matter contrasts for $\alpha_+$ of growing modes with
different $c=1,2$. }
\end{center}
\end{figure}

On the other hand, as is shown in Fig. 3, the effective EOS with
the future event  horizon has a bound of $-1 \le \omega_{\rm
m}^{\rm eff} \le -1/3$, which means that this case always provides
the accelerating universe for  $c \ge 1$. Hence we find from Fig.
5 that there is no growing mode  whenever choosing $\alpha_\pm$.
All of $\delta_{\rm FH}$ belong to decaying modes. This is
consistent with our conjecture. At this stage, we point out that
the region of $0< c <1$ with the future event horizon may lead to
the phantom phase (see Fig. 3). However, the Newtonian equations
for matter contrast do not work for $0< c <1$. Hence we could not
discuss the evolution of density perturbation in the background of
phantom phase.
\begin{figure}
\begin{center}
\includegraphics[scale=0.9]{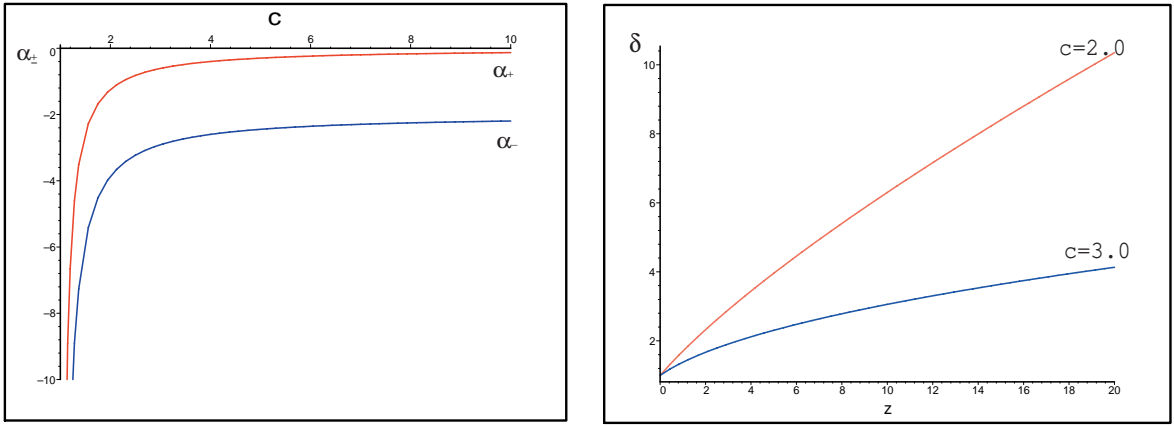}
\end{center}
\caption{\footnotesize The matter contrast as a function of the
redshift for H($t$)CDM with future event horizon (FH). The left
panel denotes the exponents $\alpha_{\pm}$  and the right one
denotes the matter contrasts for $\alpha_+$ of decaying modes with
different $c=2,3$. In the left panel, we have a forbidden region
of $0< c<1$, which may represents the phantom phase $\omega^{\rm
eff}_{\rm m}<-1$.}
\end{figure}

\section{Conclusions}
For the fixed-ratio of energy densities, we have the constant
effective EOS for the H($t$)CDM models with Hubble, particle, and
future event horizons. These indicate the different matter
productions, as a result of decaying of holographic dark energy
into CDM. In these cases, we have  definite connections for the
matter contrasts: For $\omega^{\rm eff}_{\rm m} \ge -1/3$, we have
a growing mode, while for $-1 \le \omega^{\rm eff}_{\rm m} <
-1/3$, we have no growing mode. These show clearly how the matter
contrasts evolve differently under the different matter
productions. On the other hand, for the non-interacting case of
variable ratio of energy densities, we have growing modes which
reflects  the CDM-dominated universe in the very early universe.

\section*{Acknowledgements}
H.W. Lee was in part supported by KOSEF, Astrophysical Research
Center for the Structure and Evolution of the Cosmos at Sejong
University. Y. Myung  was in part supported by  the Korea Research
Foundation (KRF-2006-311-C00249) funded by the Korea Government
(MOEHRD).


\begin{thebibliography}{}

\bibitem{Pad} V.~Sahni and A.~A.~Starobinsky,
  Int.\ J.\ Mod.\ Phys.\  D {\bf 9} (2000) 373
  [arXiv:astro-ph/9904398];

  T.~Padmanabhan,
  Phys.\ Rept.\  {\bf 380} (2003) 235
  [arXiv:hep-th/0212290];

 P.~J.~E.~Peebles and B.~Ratra,
  Rev.\ Mod.\ Phys.\  {\bf 75} (2003) 559
  [arXiv:astro-ph/0207347];

J.~Frieman, M.~Turner and D.~Huterer,
  arXiv:0803.0982 [astro-ph].

\bibitem{Ozer} M.~Ozer and M.~O.~Taha,
  Phys.\ Lett.\  B {\bf 171} (1986) 363;

   K.~Freese, F.~C.~Adams, J.~A.~Frieman and E.~Mottola,
  Nucl.\ Phys.\  B {\bf 287} (1987) 797.

\bibitem{dim} J.~S.~Alcaniz and J.~A.~S.~Lima,
  Phys.\ Rev.\  D {\bf 72} (2005) 063516
  [arXiv:astro-ph/0507372];

J.~Grande, R.~Opher, A.~Pelinson and J.~Sola,
  JCAP {\bf 0712} (2007) 007
  [arXiv:0709.2130 [gr-qc]].
\bibitem{wm}
  P.~Wang and X.~H.~Meng,
  Class.\ Quant.\ Grav.\  {\bf 22} (2005) 283
  [arXiv:astro-ph/0408495].

\bibitem{Fabris} J.~C.~Fabris, I.~L.~Shapiro and J.~Sola,
  JCAP {\bf 0702} (2007) 016
  [arXiv:gr-qc/0609017].
\bibitem{Barrow} J.~D.~Barrow and T.~Clifton,
  Phys.\ Rev.\  D {\bf 73} (2006) 103520
  [arXiv:gr-qc/0604063].

\bibitem{CKN} A.~G.~Cohen, D.~B.~Kaplan and A.~E.~Nelson,
  Phys.\ Rev.\ Lett.\  {\bf 82} (1999) 4971
  [arXiv:hep-th/9803132].


\bibitem{HSU} S.~D.~H.~Hsu,
  Phys.\ Lett.\  B {\bf 594} (2004) 13
  [arXiv:hep-th/0403052].


\bibitem{LI} M.~Li,
  Phys.\ Lett.\  B {\bf 603} (2004) 1
  [arXiv:hep-th/0403127].




\bibitem{Myung2} Y.~S.~Myung,
  Phys.\ Lett.\  B {\bf 610} (2005) 18
  [arXiv:hep-th/0412224].



\bibitem{HOV}  R.~Horvat,
  Phys.\ Rev.\  D {\bf 70} (2004) 087301
  [arXiv:astro-ph/0404204].




\bibitem{Zim3} D.~Pavon and W.~Zimdahl,
  Phys.\ Lett.\  B {\bf 628} (2005) 206
  [arXiv:gr-qc/0505020];

W. Zimdahl, W.~Zimdahl,
  Int.\ J.\ Mod.\ Phys.\  D {\bf 14} (2005) 2319
  [arXiv:gr-qc/0505056];
J.~Valiviita, E.~Majerotto and R.~Maartens,
  arXiv:0804.0232 [astro-ph].


\bibitem{KLM}
  H.~Kim, H.~W.~Lee and Y.~S.~Myung,
  Phys.\ Lett.\  B {\bf 632} (2006) 605
  [arXiv:gr-qc/0509040].

\bibitem{WGA} B.~Wang, Y.~g.~Gong and E.~Abdalla,
  Phys.\ Lett.\  B {\bf 624} (2005) 141
  [arXiv:hep-th/0506069].

 \bibitem{Myungs}
  Y.~S.~Myung,
  Phys.\ Lett.\  B {\bf 626} (2005) 1
  [arXiv:hep-th/0502128].

  \bibitem{KLMM}
  K.~Y.~Kim, H.~W.~Lee and Y.~S.~Myung,
  Mod.\ Phys.\ Lett.\  A {\bf 22} (2007) 2631
  [arXiv:0706.2444 [gr-qc]].

\bibitem{MFB}
  V.~F.~Mukhanov, H.~A.~Feldman and R.~H.~Brandenberger,
  Phys.\ Rept.\  {\bf 215} (1992) 203.

\bibitem{BRAN}
  R.~H.~Brandenberger,
  Lect.\ Notes Phys.\  {\bf 646} (2004) 127
  [arXiv:hep-th/0306071].



\bibitem{Waga}
  R.~C.~Arcuri and I.~Waga,
  Phys.\ Rev.\  D {\bf 50} (1994) 2928.

  \bibitem{Borges}
  H.~A.~Borges, S.~Carneiro, J.~C.~Fabris and C.~Pigozzo,
  Phys.\ Rev.\  D {\bf 77} (2008) 043513
  [arXiv:0711.2689 [astro-ph]].

















\end{thebibliography}
\end{document}